\documentclass[%
 aip,
 amsmath,amssymb,
 reprint,%
]{revtex4-1}

\usepackage{graphicx}
\usepackage{dcolumn}
\usepackage{bm}

\usepackage[utf8]{inputenc}
\usepackage[T1]{fontenc}
\usepackage{mathptmx}
\usepackage{etoolbox}
\usepackage{comment}
\usepackage{acronym}

\makeatletter
\def\@email#1#2{%
 \endgroup
 \patchcmd{\titleblock@produce}
  {\frontmatter@RRAPformat}
  {\frontmatter@RRAPformat{\produce@RRAP{*#1\href{mailto:#2}{#2}}}\frontmatter@RRAPformat}
  {}{}
}%

\makeatother

\begin{document}

\preprint{AIP/123-QED}

\title[Current pulse generator: A circuit for programming RRAM in current mode]{Current pulse generator: A circuit for programming RRAM in current mode}

\author{B. Zhang}
\author{P. Gibertini}%

\author{M. Akbari}

\author{E. Covi}
\affiliation{ 
Zernike Institute of Advanced Materials \& CogniGron, University of Groningen\\
Nijenborgh 3, 9747 AG Groningen, Netherlands
}%

\date{\today}

\begin{abstract}
Switching uniformity, as a major challenge, hinders the practical implementation of \ac{RRAM} in memory application. Operating \ac{RRAM} in current mode, is proposed as an efficient method to improve programming schemes accuracy within the finite readout window. In this article, we demonstrate a current generator circuit to perform current programming on \ac{RRAM}. Current mirror topology is used in our circuit to convert an external pulse voltage into a pulse current fed to \ac{RRAM} directly with an amplitude equivalent with the DC reference current. The targeting ranges of \ac{RRAM}'s programming current are up to 400\,\textmu A and, in that case, our proposed circuit achieved minimum current mismatch of 1\%.
\end{abstract}

\maketitle

\acrodef{IMC}[IMC]{In memory computing}
\acrodef{CMOS}[CMOS]{complementary metal–oxide–semiconductor}
\acrodef{BEOL}[BEOL]{back end of the line}
\acrodef{NVM}[NVM]{non-volatile memory}
\acrodef{RRAM}[RRAM]{Resistive random access memory}
\acrodef{eNVM}[eNVM]{embedded nonvolatile memory}
\acrodef{IC}[IC]{integrated circuit}
\acrodef{1T1R}[1T1R]{one-transistor-one-resistor}
\acrodef{ISPVA}[ISPVA]{incremental step pulse with verify algorithm}
\acrodef{IGVVA}[IGVVA]{incremental gate voltage with verify algorithm}
\acrodef{PVA}[PVA]{program-verify algorithm}
\acrodef{MOS}[MOS]{metal-oxide-semiconductor}
\acrodef{HRS}[HRS]{high resistive state}
\acrodef{LRS}[LRS]{low resistive states}
\acrodef{2M1R1B}[2M1R1B]{two current mirrors - one RRAM - one voltage buffer}
\acrodef{NMOS}[NMOS]{n-type metal-oxide-semiconductor}
\acrodef{PMOS}[PMOS]{p-type metal-oxide-semiconductor}
\acrodef{W/L}[W/L]{width over length}
\acrodef{VMM}[VMM]{vector matrix multiplication}
\acrodef{ADC}[ADC]{analog-to-digital converter}

\section{\label{sec:level1}Introduction}

\ac{RRAM} technology has received widespread attention for the potential as a promising future \ac{NVM}. Its non-volatile switching behavior ensures good retention even at elevated temperatures\cite{suga2016highly}, while program/erase cycling shows strong endurance \cite{yang2010high}, and simple device structure enables high-density integration\cite{seok2014review, 11349411}, making \ac{RRAM} an attractive solution for \ac{eNVM}. The \ac{RRAM} properties, such as nonvolatile behavior, multilevel operation, and good scaling capability, align closely also with the requirements of various computing applications in image classification and voice recognition \cite{wan2022compute}. Despite its promising characteristics, \ac{RRAM} technology still faces critical challenges in variability, which hinder its adoption as a mature embedded nonvolatile memory technology. While material and device engineering can partially address this issue by tuning oxygen non-stoichiometry of the switching layer \cite{skaja2018reduction} to stabilize the filament dynamic or by doping the oxide to decrease the grain size \cite{lee2015effect} and suppress the later diffusion \cite{he2024impact}, the programming scheme also directly influences the evolution dynamics of conductive filaments and control switching behavior. Properly designed programming schemes can suppress abrupt switching behavior, reduce stochastic variations, and improve both switching uniformity \cite{lim2025impact}. 

Generally, two operation modes of circuits are distinguished: voltage mode and current mode. In voltage mode, the programming voltage pulse is typically directly applied to the two-terminal \ac{RRAM} device to modulate its resistance and achieve the target states. However, particularly during the SET operation, where a conductive filament changes the \ac{RRAM} resistive state from high to low, to prevent the device from the potential damage caused by the abrupt increase of current, a current compliance is usually used to limit the maximum SET current. In two-terminal devices, this is typically achieved by external tools or passive elements such as resistors. With the development of two terminal \ac{RRAM} cell to three terminal \ac{1T1R} structure, the external gate terminal of MOS transistor brings more possibility of \ac{RRAM} programming in voltage mode, as the current compliance is obtained by applying a voltage to the gate of the transistor, which acts as a voltage controlled current source. Advanced programming schemes as \ac{ISPVA} \cite{perez2021optimization}, \ac{IGVVA} \cite{milo2021optimized} and dispersion aware \ac{PVA} \cite{7177062} are introduced into \ac{RRAM} operation to optimize \ac{RRAM} performance in multilevel programming for both information writing and storage as \ac{eNVM}. It has been demonstrated that these programming schemes are capable of controlling the programmed conductance precisely. However, it should be noted that the verification of accurate programming with incremental pulse and conductance states can result in an increase in power consumption.\par

As for current mode, the current programming pulse \cite{8662393} (or sweep \cite{wt2011improved}) is generated by reference current source and directly fed to the \ac{RRAM} device. Operating \ac{RRAM} devices with current is quite promising to improve device performance in reliable resistance switching or better uniformity of \ac{LRS} distribution \cite{5488912}. From the circuit design point of view, the reference current source could drive a broader range of loads in current mode, hence the switching process could be less influenced by the resistive change of \ac{RRAM} and the related change of transistor working point \cite{10461986}. Moreover, there is no need for setting an additional current compliance during SET operation, since the current is directly fed to the \ac{RRAM}.\par

In this work, we propose a \ac{2M1R1B} circuit to execute current programming of \ac{RRAM} devices. By utilizing pairs of current mirrors to copy the current from a reference DC current source and chop transistors to shape the programming pulse, our proposed circuit aims at feeding well defined current programming pulses to \ac{RRAM} devices. With DC measurements on SET and RESET branches in our circuits, we measure the mirror factor at various current amplitudes to evaluate its deviation from target factor 1.0 and the operation range by varying DC voltage supply of two branches. Moreover, mismatch of MOSFETs is characterized by evaluating the mirror factor deviation across the whole wafer. With transient characterization, we evaluate the mirror factors with pulses to assess the stability of mirroring action. Furthermore, by tuning the chop pulses' edge we characterized the mirror performance of our circuit at different time scale. 
Finally, we discuss the non idealities and possible improvement of our circuit to optimize current programming of \ac{RRAM}.\par   


\section{Description of the circuit}

\begin{figure}
    \centering
    \includegraphics[width=1.0\columnwidth]{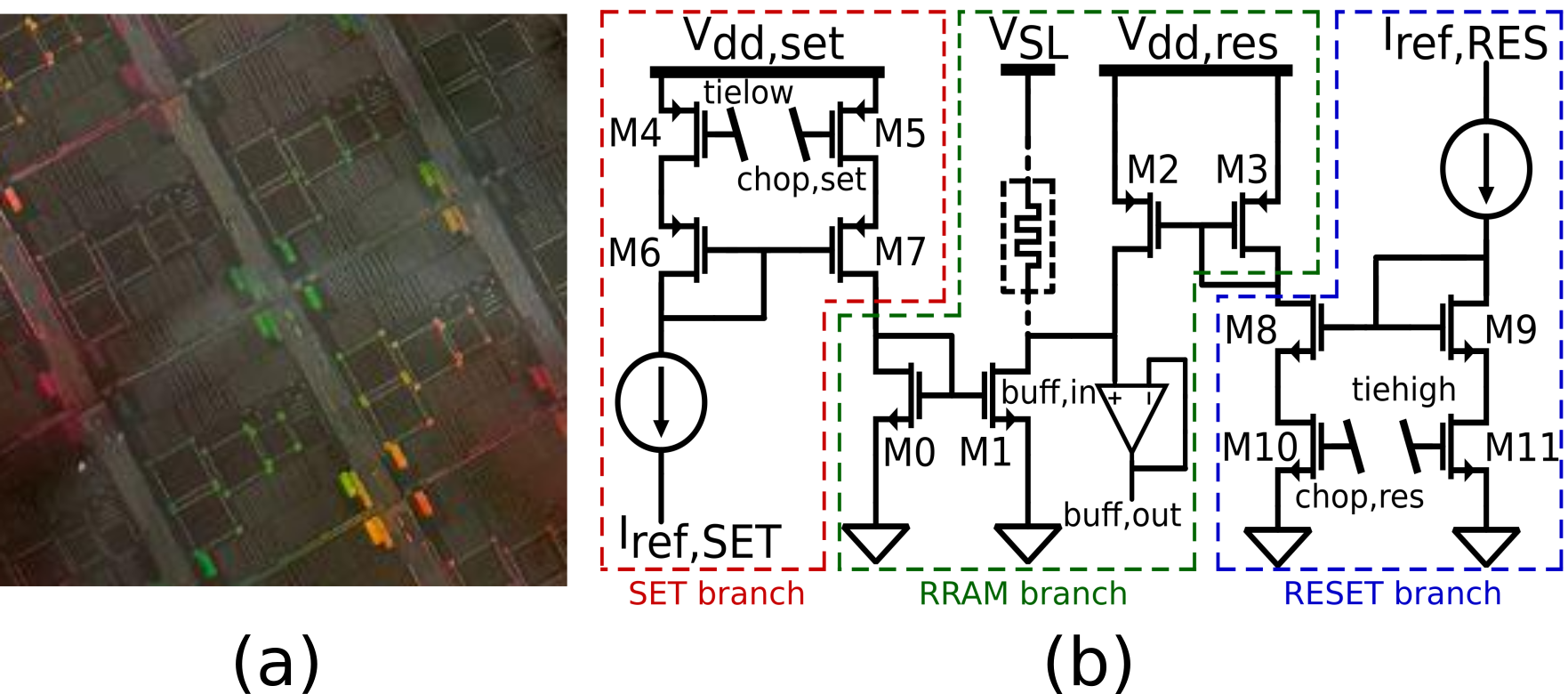}
    \caption{(a) Photograph of the fabricated wafer containing multiple dies. (b) Schematic of the \ac{2M1R1B} circuit.}
    \label{fig:wafer and circuit}
\end{figure}

The proposed circuit, designed in a standard 180\,nm technology node, is shown in Fig.~\ref{fig:wafer and circuit}. Figure~\ref{fig:wafer and circuit}a shows the taped-out wafer that includes our circuit, whereas its schematic is sketched in Fig.~\ref{fig:wafer and circuit}b. The circuit is divided in three main branches: SET branch, RESET branch, and \ac{RRAM} branch. The set branch, shown in red dashed line, generates current pulses to switch the \ac{RRAM} to its SET state. Two \ac{PMOS} devices, M4 and M5, are used as chopping switches to convert a DC current, $I_{\mathrm{ref,set}}$, into a current pulse. The gate terminal of M4 is always grounded to provide a large voltage, $V_{\mathrm{dd,set}}$, across its source-gate voltage and keep M4 in the deep-triode region with a $V_{\mathrm{SD,M4}}$=0\,V. Therefore, when a negative large voltage pulse, from $V_{\mathrm{dd,set}}$ to 0\,V, is applied to the gate of the counterpart device, M5, the source-drain voltage of M5, $V_{\mathrm{SD,M5}}$, will immediately tend to zero to keep M5 in the deep-triode region for the duration of the input voltage pulse. In case of ideal symmetry, both M4 and M5 will show the same small resistances leading to a high matching between the devices of the p-channel current mirror, M6:M7, which increases the accuracy of the mirroring process. Next, as long as the input voltage pulse is active, a current as large as the reference current, $I_{\mathrm{ref,set}}$, is mirrored into the output of the set circuit, M0:M1, producing a current pulse for the duration of the input voltage pulse. The generated negative current pulse mirroring into the \ac{RRAM} branch, shown in green dashed line, switches the \ac{RRAM}, and the voltage across the \ac{RRAM} is measured by a voltage buffer. 

Regarding the RESET circuit, shown in blue dashed line, a similar strategy is employed by which the \ac{RRAM} can be reset in current mode. Briefly, the gate of M11 is always connected to additional DC supply, $V_{\mathrm{dd}}$ to keep M11 in deep-triode region and as soon as a positive voltage pulse, from 0\,V to $V_{\mathrm{dd}}$, is applied to the gate of its counterpart device, M10, the DC reference current ($I_{\mathrm{ref,res}}$) is mirrored into the output of the reset circuit by current mirror M9:M8. Thus, for duration of the input voltage pulse, a current pulse with an amplitude of $I_{\mathrm{ref,res}}$ is generated by which the \ac{RRAM} is sourced through current mirror M3:M2. In this work, voltage sensing is used to read the exact resistance of the \ac{RRAM} by draining read current pulse to \ac{RRAM} to evaluate its \ac{LRS} or \ac{HRS} status.\par

\section{Results}
\subsection{\label{sec:level2}DC characterization}

To verify the circuit’s performance, DC measurements are performed on SET as well as RESET branches first.

\begin{figure}
    \centering
    \includegraphics[width=0.75\columnwidth]{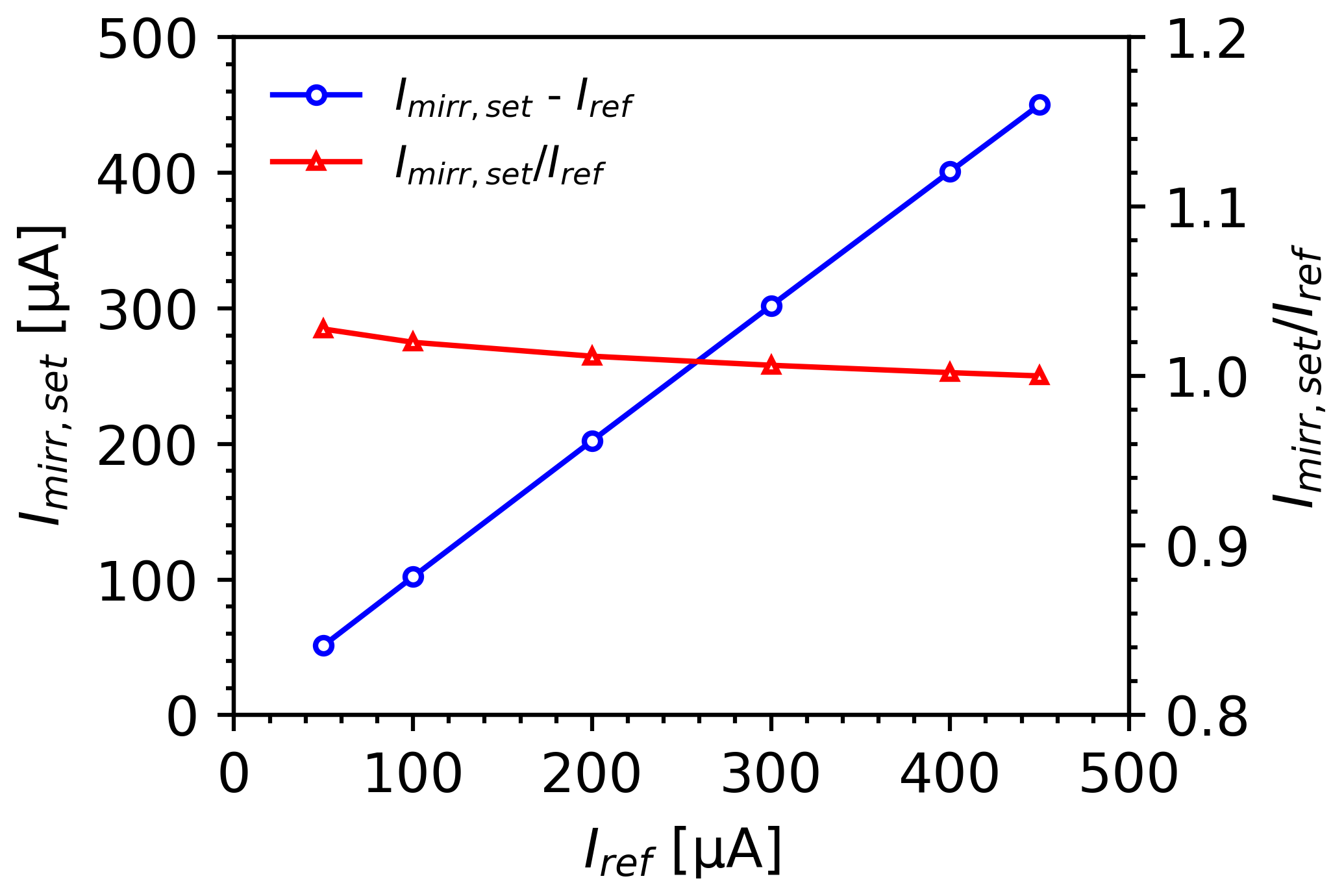}
    \caption{DC characterization of the mirror in the SET branch. The mirrored current $I_{\mathrm{mirr,set}}$ in branch M5/M7/M0 increases linearly with $I_{\mathrm{ref,set}}$ varying from 50\,\textmu A to 450\,\textmu A (blue curve). The corresponding mirror factor across the whole $I_{\mathrm{ref,set}}$ is kept approximately at 1.0 (red curve) with a maximum aberration of 3\% at 50\,\textmu A.} 
    \label{fig:dc SET fa}
\end{figure}

In the DC measurements on the SET branch, the reference current $I_{\mathrm{ref,set}}$ is programmed from 50\,\textmu A up to 450\,\textmu A as DC input of the \ac{PMOS} mirror (M6:M7). The power supply $V_{\mathrm{dd,set}}$ is set at 5\,V, and the mirrored current is also read from same terminal by subtracting the reference current. Moreover, the second power supply is used to provide common ground for the circuit. The corresponding mirrored current and the mirror factor are shown in Fig.~\ref{fig:dc SET fa}. As the reference current is increased, the mirrored current in branch M5/M7/M0 exhibits a linear relationship. The mirror factor is typically maintained at approximately 1.0 across the entire current range. Nonetheless, a departure from the optimal ratio of 1.0 persists in the lower current range, with a maximum aberration of 3\% observed at 50\,\textmu A. This current mismatch could be ascribed to the voltage mismatch of M6:M7. In SET branch, M4 and M5 are in the deep triode region during the operation to provide a matching of $V_{\mathrm{s,M6}}$ and $V_{\mathrm{s,M7}}$. But the node $V_{\mathrm{d,M7}}$ is affected by the $V_{\mathrm{dd,set}}$, while $V_{\mathrm{d,M6}}$ is regulated by current source. While operating small current with a high voltage supply of $V_{\mathrm{dd,set}}$, the node $V_{\mathrm{d,M7}}$ could be pulled up to higher voltage by $V_{\mathrm{dd,set}}$, which leads to a higher mirroring current.\par

\begin{figure}
    \includegraphics[width=0.75\columnwidth]{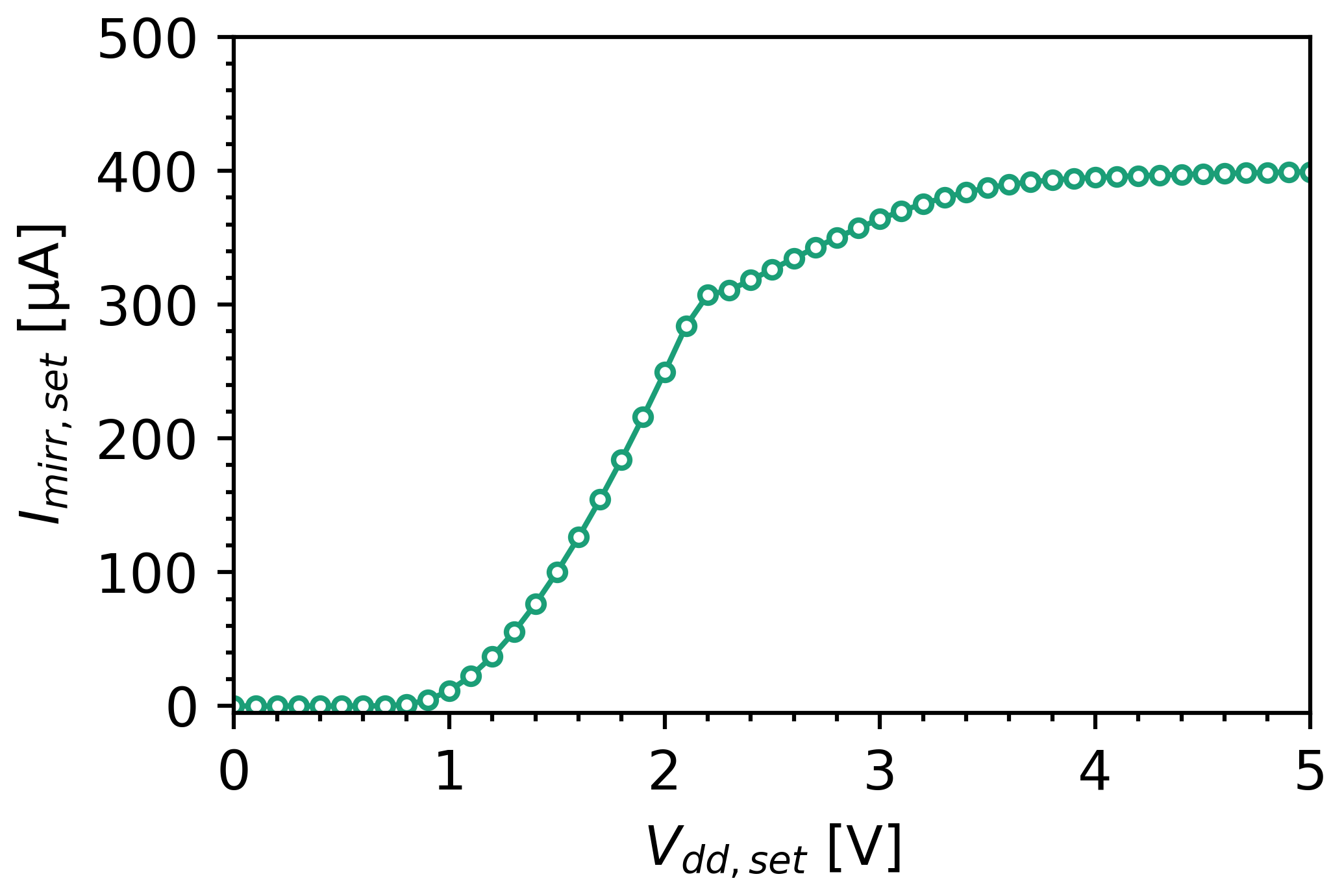}
    \caption{Operation range assessment of SET branch with DC measurement. $I_{\mathrm{mirr,set}}$ is measured while sweeping $V_{\mathrm{dd,set}}$ from 0\,V to 5\,V with $V_{\mathrm{chop,set}}$ at 0\,V.}
    \label{fig:dc SET op}
\end{figure}

After characterizing the mirror factor, we assess the operating range of our circuit by sweeping $V_{\mathrm{dd,set}}$ from 0 to 5\,V. Same as previous measurement, the mirrored current is read from the power supply by subtracting the reference current. The voltage level of the $V_{\mathrm{chop,set}}$ is established 0\,V to maintain a matching of M4:M5. The reference current has been set to 400\,\textmu A. The relationship between the measured current and the $V_{\mathrm{dd,set}}$ is illustrated in Fig.~\ref{fig:dc SET op}. The minimum $V_{\mathrm{dd,set}}$ required by the SET branch is approximately 3.5\,V to mirror the reference DC current. With the further decreasing of $V_{\mathrm{dd,set}}$ to 2\,V, the mirror current drops gradually, which is caused by the voltage mismatch of $V_{\mathrm{sd}}$ of M6:M7. Once the $V_{\mathrm{dd,set}}$ is lower than 2\,V, the current source can not maintain the DC current output under low voltage supply, which results in a fast current drop to zero.

\begin{figure}
    \centering
    \includegraphics[width=0.75\columnwidth]{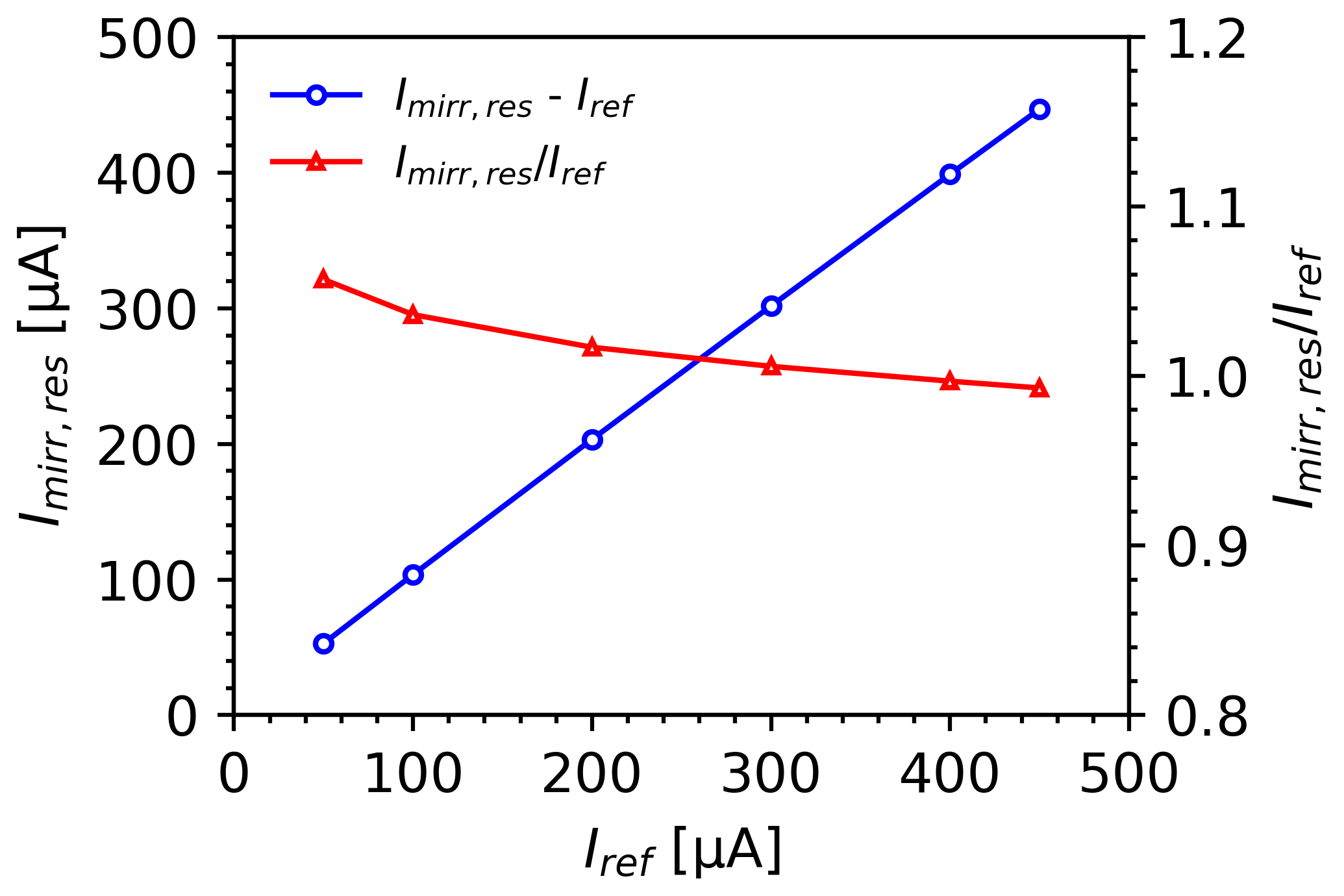}
    \caption{DC characterization of the mirror in the RESET branch. The mirrored current $I_{\mathrm{mirr,res}}$ in branch M3/M8/M10 increases linearly with $I_{\mathrm{ref,res}}$ varying from 50\,\textmu A to 450\,\textmu A (blue curve). The corresponding mirror factor across the whole $I_{\mathrm{ref,res}}$ is kept approximately at 1.0 (red curve) with a maximum aberration 6\% at 50\,\textmu A.}
    \label{fig:dc RESET fa}
\end{figure}

As in the case with DC measurements performed on the SET branch, the mirror factor and operation range of $V_{\mathrm{dd,res}}$ are characterized on the RESET branch. The reference current $I_{\mathrm{ref,set}}$ is programmed from 50\,\textmu A up to 450\,\textmu A as DC input of the \ac{NMOS} mirror (M8:M9). The power supply $V_{\mathrm{dd,res}}$ is set to 5\,V, and the mirrored current is also read from the power supply terminal. Same as the prior measurement configuration, the shared ground is furnished by the extra power supply. The mirror factor is characterized in the first instance. The corresponding mirrored current and the mirror factor are displayed in Fig.~\ref{fig:dc RESET fa}. The mirror current and the reference current exhibit a linear relationship. The mirror factor is mainly maintained at approximately 1.0 across the whole operation current range. As was the case in the SET branch, in the lower current range there is a deviation observed, with a maximum 6\% at 50\,\textmu A. As discussed in the previous measurement, the voltage mismatch of M8:M9 could cause this deviation while operating in the low current range. In RESET branch, M10 and M11 are in the deep triode region during the operation to provide a matching of $V_{\mathrm{s,M8}}$ and $V_{\mathrm{s,M9}}$. Since all transistors are \ac{NMOS}, $V_{\mathrm{s,M8}}$ and $V_{\mathrm{s,M9}}$ are actually pulled down to zero. In this case, the $V_{\mathrm{ds,M8}}$ is more sensitive to the mismatch of $V_{\mathrm{d,M8}}$, especially when this branch is operated with lower current. \par

\begin{figure}
    \centering
    \includegraphics[width=0.75\columnwidth]{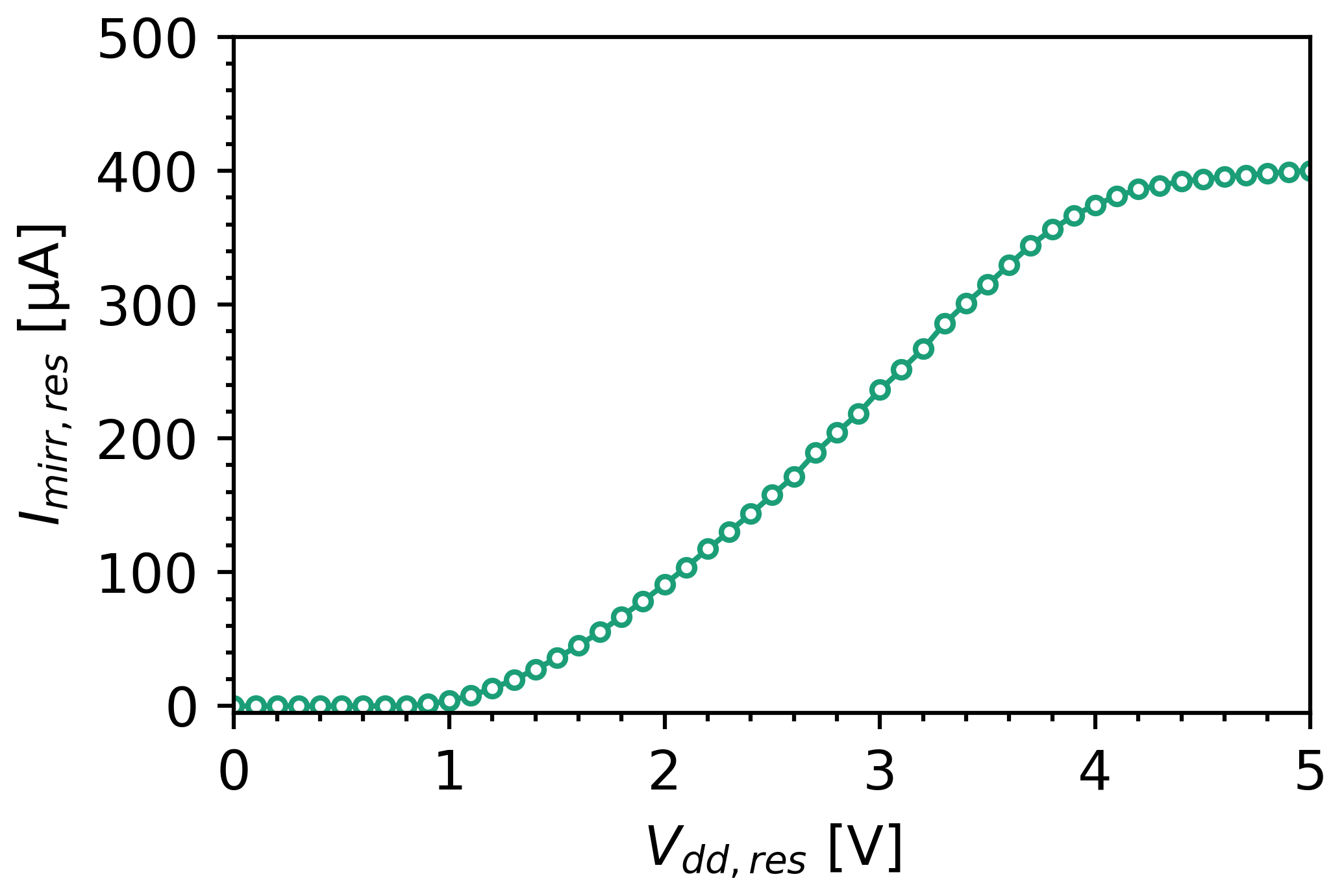}
    \caption{Operation range assessment of RESET branch with DC measurement. $I_{\mathrm{mirr,res}}$ is measured while sweeping $V_{\mathrm{dd,res}}$ from 0\,V to 5\,V with $V_{\mathrm{chop,res}}$ at 5\,V.}
    \label{fig:dc RESET op}
\end{figure}

Following the mirror factor characterization, the characterization of operation range of $V_{\mathrm{dd,res}}$ is performed on the RESET branch. The power supply is swept from 0 to 5\,V while concurrently measuring the current. The $V_{\mathrm{chop,res}}$ is set to 5\,V to maintain a high matching of M10:M11. The reference current is still set to 400\,\textmu A. In Figure~\ref{fig:dc RESET op}, the relationship between the current and $V_{\mathrm{dd,res}}$ is demonstrated. The measured current is the precise value of the mirrored current in branch M3/M8/M10. According to the $V_{\mathrm{chop,res}}$ values, the voltage supply can be lowered down to 4\,V without affecting the normal operation of the circuit. A comparison of the two branches reveals that the operation range of general voltage supply in the RESET branch is narrower than the SET branch. The origin of this minimum voltage supply difference could be the diode connected transistors M0 and M3. As introduced in previous sections, transistors M4 and M5 as well as M10 and M11 are all operated in deep triode region. In the case of ideal matching, they can pull the source voltage of mirror M6:M7 as well as M8:M9 to power supply and ground. For these two pairs of current mirrors, the diode connected gate voltages are controlled by the current source. In this configuration the diode connected \ac{NMOS} M0 and \ac{PMOS} M3 are affected by the power supply. However, for p-channel transistor, due to the lower carrier mobility, it required larger gate source voltage to carry the same amount current as n-channel transistor. Furthermore, in the diode connected configuration, the gate source voltage is equal to drain source voltage. Thus, the diode connected M3 needs a larger drain-source voltage than M0.\par

\begin{figure}
    \centering
    \includegraphics[width=0.5\textwidth]{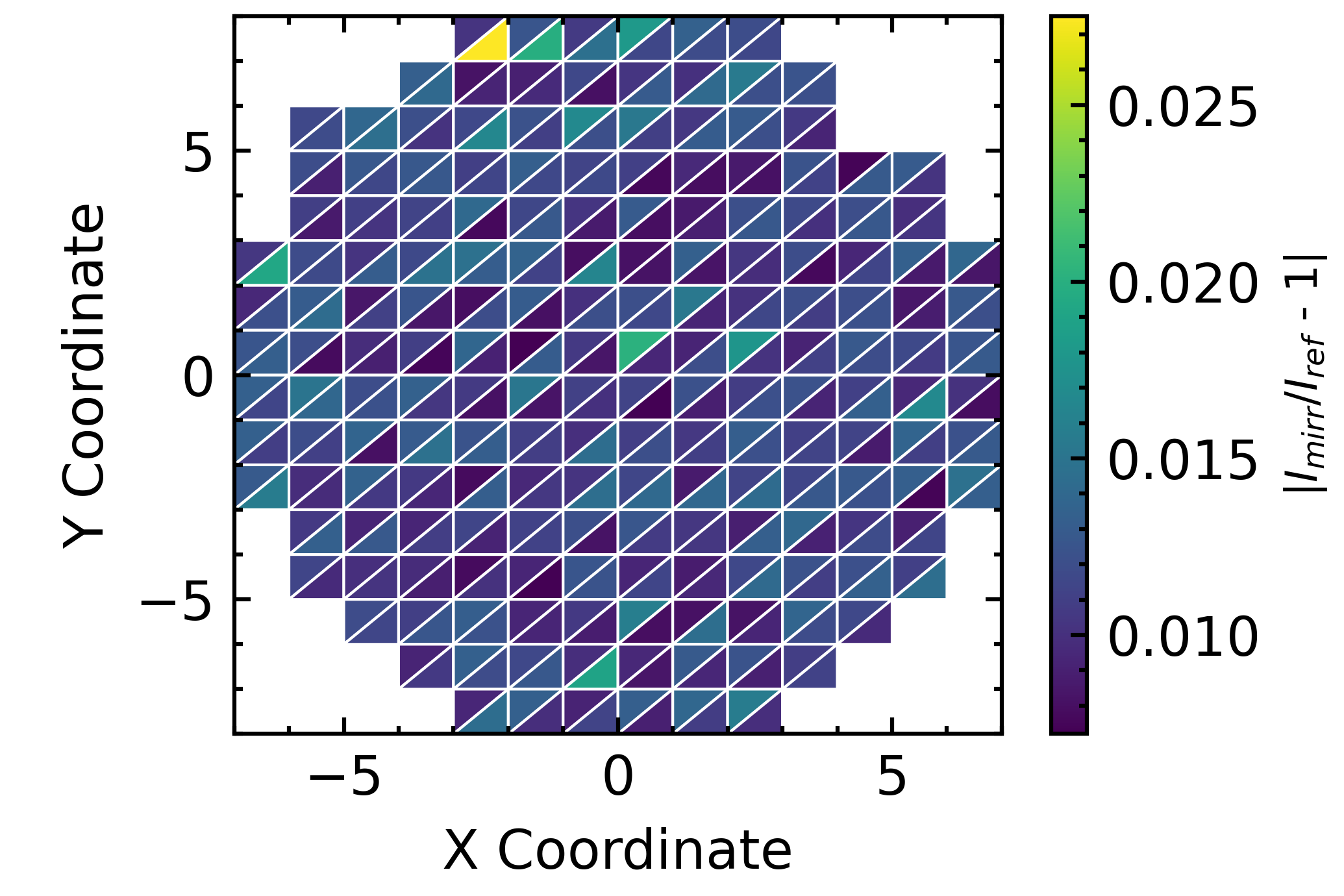}  
    \caption{\ac{CMOS} mismatch evaluation of SET branch at wafer level with DC measurement on two same \ac{2M1R1B} structures per die.}
    \label{fig:dc wafer_a}
\end{figure}

\begin{figure}
    \centering
    \includegraphics[width=0.5\textwidth]{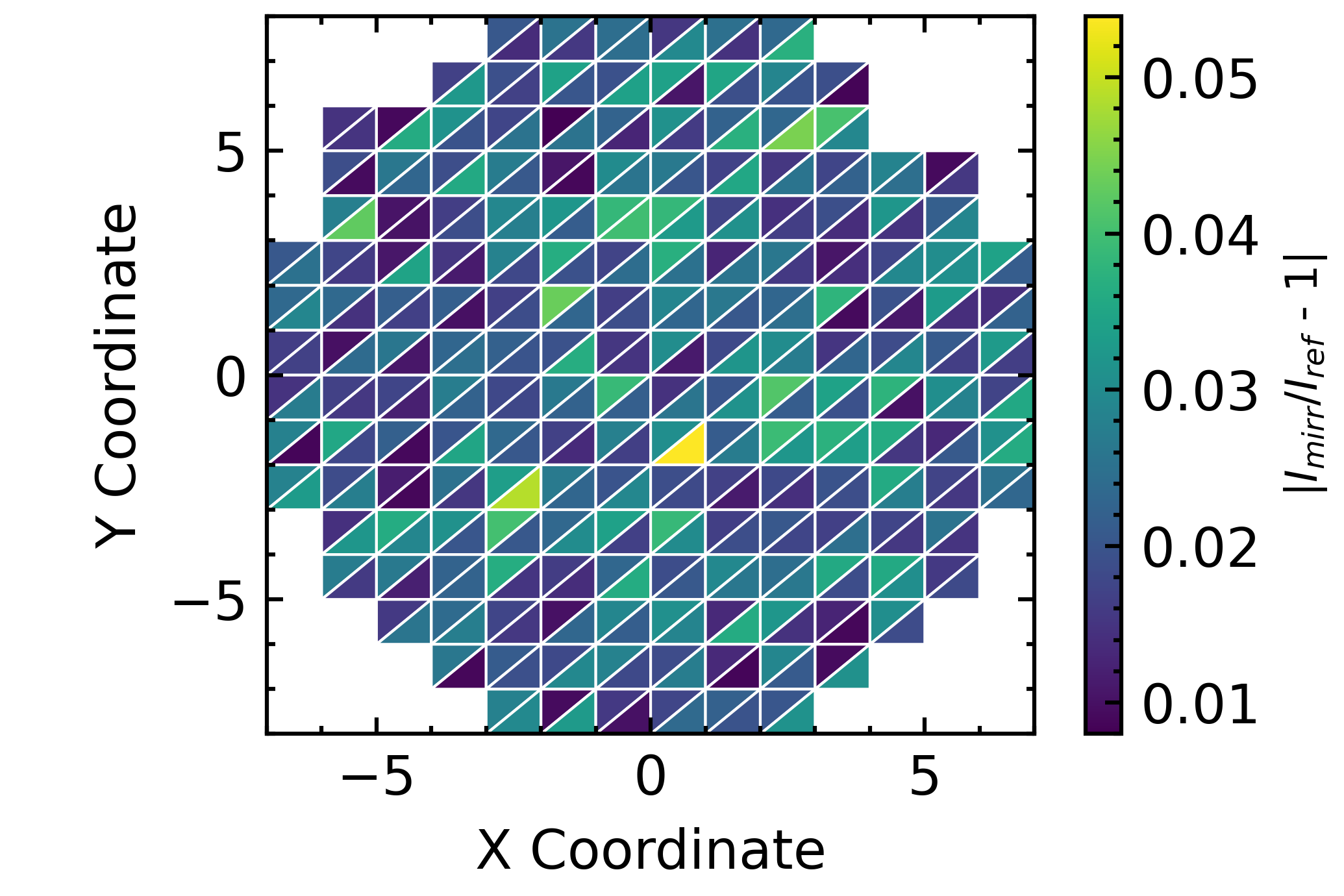}  
    \caption{\ac{CMOS} mismatch evaluation of RESET branch at wafer level with DC measurement on two same \ac{2M1R1B} structures per die.}
    \label{fig:dc wafer_b}
\end{figure}

To evaluate the impact of process variation as well as \ac{CMOS} mismatch on uniformity of average mirroring accuracy on wafer scale, DC measurements are performed on 180 dies of the wafer. For each die, two \ac{2M1R1B} circuits at different position of the die are measured with four reference current from 100\,\textmu A to 400\,\textmu A. And later the mean mirror factor of these four current values of each test circuit on each die is calculated. The average mirroring accuracy of both branches at wafer level are shown in Fig.~\ref{fig:dc wafer_a} and Fig.~\ref{fig:dc wafer_b} in the form of heatmap. The SET branch shows a homogeneous distribution of mirror factor deviation in the range of 1\% to 2\%, besides one circuit on the first intensively tested die. As for the RESET branch, the mirror factor deviation is in the range of 1\% to 4\% with several \ac{2M1R1B} structures from different dies show highest deviation in the middle of wafer. Since the RESET process of \ac{RRAM} operation has a higher robustness on current than SET operation, the deviation is acceptable. Compared to the RESET branch, the average mirroring accuracy of SET branch is less influenced by process variation and \ac{CMOS} mismatch, which is crucial to program \ac{RRAM} to different \ac{LRS} states with precise current control during the operation. As for RESET branch, even with deviation up to 4\%, it is still able to mirror enough current to reset the device.\par

\subsection{\label{sec:level2}Transient measurement}

To characterize the circuit’s performance under the pulse operation, transient measurements are performed on SET as well as RESET branches.\par

The performance of SET branch in transient operation is firstly evaluated by executing following set of measurements on silicon. With varying the reference current value, the mirror factor can be calculated by reading the amplitude of corresponding current pulses. Furthermore, with set of different rise time of the chop pulse, the edge of current pulse can be verified. The mirror factor of \ac{PMOS} current mirror M6:M7 is firstly measured. In this case, the current source is programmed to generated DC current from 50\,\textmu A up to 400\,\textmu A. As mentioned perviously in Section~2, a negative large voltage pulse, from $V_{\mathrm{dd,set}}$ to 0, is applied to the gate of M5 in a waveform with period 20\,\textmu s and rise time 1\,\textmu s as well as duration time 10\,\textmu s. The corresponding $V_{\mathrm{dd,set}}$ is setup to DC 5\,V and later changed to 4\,V as in DC measurement to evaluate the mirror properties under lower operation voltage. After all input signal setup, the mirror current is collected at source of M0 and measured by oscilloscope with an input resistance 50\,$\mathrm{\Omega}$. Each different measurement is repeated 50 times and averaged to minimize the influence of the noise coming from the setup on the measurements.\par

\begin{figure}
    \centering
    \includegraphics[width=0.75\columnwidth]{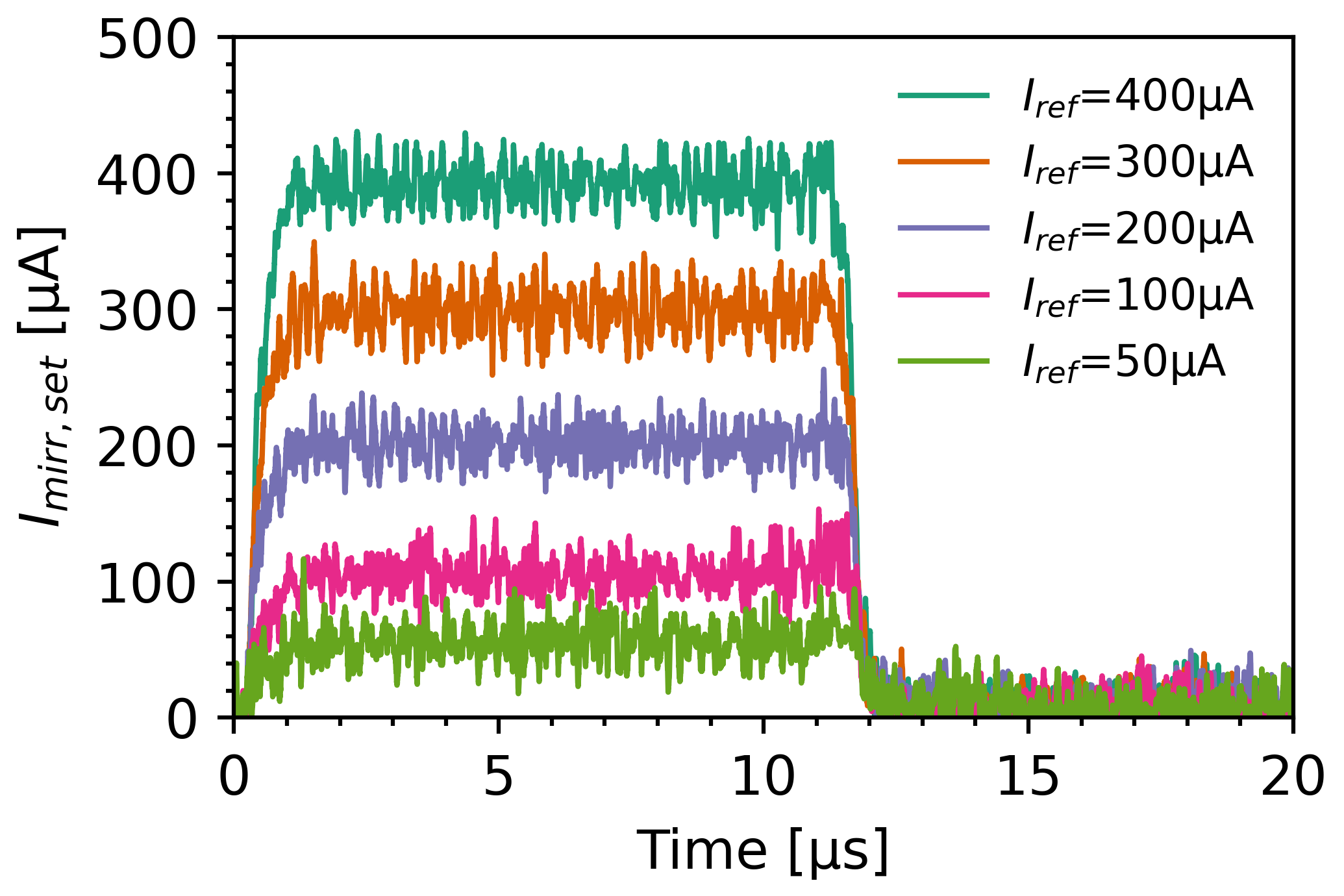}
    \caption{The average of 50 mirrored current pulses with $I_{\mathrm{ref}}$ ranging from 50\,\textmu A to 400\,\textmu A measured from transient measurements on the SET branch.}
    \label{fig:tran SET}
\end{figure}

\begin{figure}
    \centering
    \includegraphics[width=0.75\columnwidth]{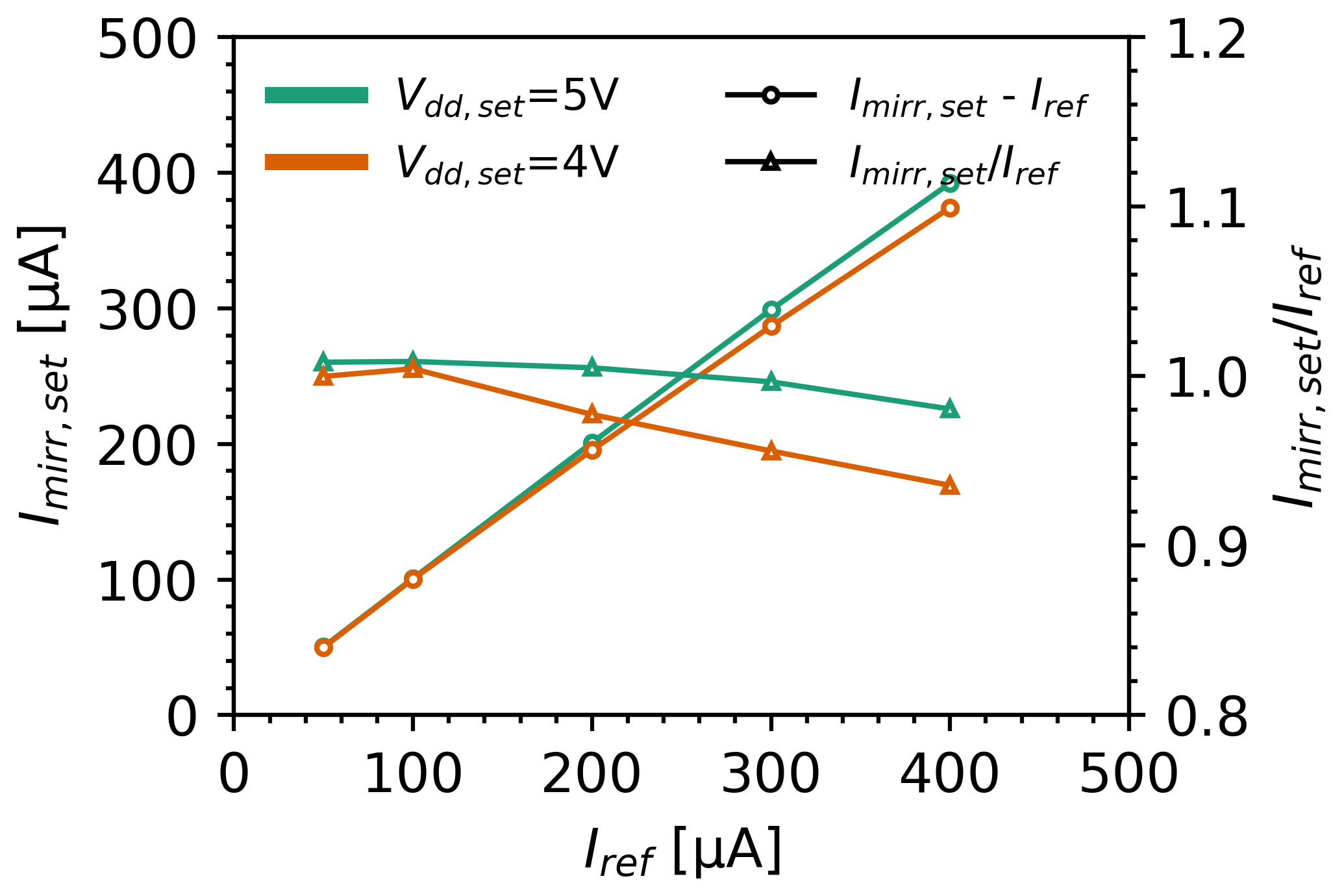}
    \caption{$I_{\mathrm{mirr,set}}$ - $I_{\mathrm{ref}}$ relation and corresponding mirror factors with $I_{\mathrm{ref}}$ ranging from 50\,\textmu A to 400\,\textmu A under variable $V_{\mathrm{dd,set}}$.}
    \label{fig:tran SET analyse fa}
\end{figure}

The characterization results are shown in Fig.~\ref{fig:tran SET}. All current pulses are characterized with sharp pulse edge with 1\,\textmu s rise time and 10\,\textmu s duration. The current pulse waveform basically matches our chop pulse's waveform. At the end of pulse rising edge, the slope of the pulse edge becomes gradual, which indicates a well current controlling of our circuit. In the duration region, although the signal is noisy, the average of amplitude of each current pulse basically match its reference value from 50\,\textmu A to 400\,\textmu A, which basically covers our target \ac{RRAM} operation current range. To further evaluate the \ac{PMOS} current mirror’s functionality, we compare the mean value of $I_{\mathrm{mirr,set}}$ with $I_{\mathrm{ref}}$ and further calculate the mirror factor as shown in Fig.~\ref{fig:tran SET analyse fa}. Same as DC characterization, there is a clear quasi linear relation between $I_{\mathrm{mirr,set}}$ and $I_{\mathrm{ref}}$ and the overall mirroring factor in reference current range (50\,\textmu A to 400\,\textmu A) is mainly maintained at approximately 1.0 across the whole operation current range with a maximum of 2\% at 400\,\textmu A. Both current pulses' waveform in Fig.~\ref{fig:tran SET} and the less deviated average mirror factor indicate that the SET branch is able to convert the DC reference to pulse by chopping operation and there is no impactful parasitic capacitance observed in the converted current pulse, which is usually characterized by spikes at the end of pulse rising edge. To exam the mirroring accuracy under lower power consumption, $V_{\mathrm{dd,set}}$ is lower to 4\,V. As the orange curve shown in Fig.~\ref{fig:tran SET analyse fa} the mirror current drop in the range of 300\,\textmu A to 400\,\textmu A with a maximum 7\% at 400\,\textmu A. However, in the lower current range from 50\,\textmu A to 200\,\textmu A, the mirror factor shows a small deviation less than 3\%, which brings possibility to program \ac{RRAM} with small current to further lower the power consumption.\par

\begin{figure}
    \centering
    \includegraphics[width=0.75\columnwidth]{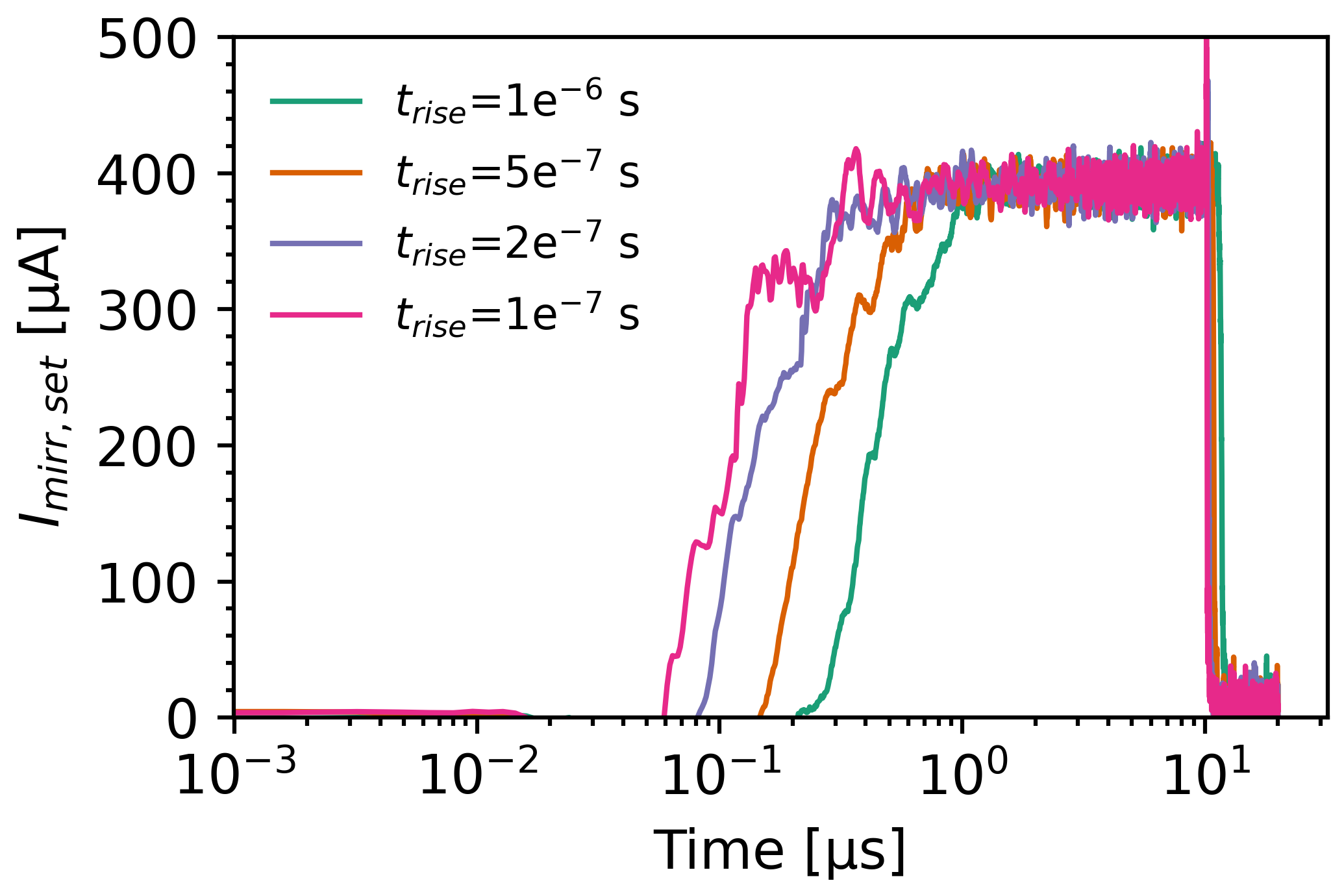}
    \caption{The average of 50 mirrored current pulses with an $I_{\mathrm{ref}}$ of 400\,\textmu A and variable chop pulse edges ranging from 100\,ns to 1\,\textmu s obtained from transient measurements on the SET branch. The time axis is on a log scale.}
    \label{fig:tran SET analyse dy}
\end{figure}

Additionally, the SET branch as a current pulse generator, its ability to generate a current pulse with a waveform defined by the $V_{\mathrm{chop,set}}$ is important to the further \ac{RRAM} programming, where the filament evolution during SET is influenced by the current. To do this, the chop pulse with rise time from 100\,ns to 1\,\textmu s, meanwhile keep the duration time at 10\,\textmu s and period at 20\,\textmu s. At the mean while, $V_{\mathrm{dd,set}}$ is set to 5\,V to provide enough power supply to drive the current. The final measurements’ results are shown in Fig.~\ref{fig:tran SET analyse dy}. To further observe and compare current pulse edge, the time scale is converted to log scale. The edge of each current pulse is in the corresponding time scale from 200\,ns up to 1\,\textmu s, which is comparable to the chop pulse. And later this current pulse will be further mirrored to the \ac{RRAM} branch to SET the device, which aims at studying the SET dynamic of \ac{RRAM} in current programming mode. However, in the case of rise time at 100\,ns, there is a small platform observed at the rising edge. One of the possible explanation is that the signal is reflected in the waveguide. When the circuit is operated with maximum frequency 10\,MHz, the probe as well as cables behave as waveguide. All cables have impedance of 50\,$\mathrm{\Omega}$, but the metal pads are not exactly same as cables. As a result of this impedance mismatch, the signal could be reflected in the waveguide.\par

The performance of RESET branch in transient operation is further evaluated with set of measurements executed on silicon. The mirror current value is mainly discussed in this part, since the amount of current it can drive is the main concern for RESET operation of \ac{RRAM}. Same as SET transient measurement, the current source is programmed to generated DC current from 50\,\textmu A up to 400\,\textmu A. Two power supplies are used to set $V_{\mathrm{dd,res}}$ to 5\,V and common ground for the circuit. $V_{\mathrm{chop,res}}$ is programmed to positive large voltage pulse, from 0 to $V_{\mathrm{dd,res}}$, is applied to the gate of M10 in a waveform with period 20\,\textmu s and rise time 1\,\textmu s as well as duration time 10\,\textmu s. \par

\begin{figure}
    \centering
    \includegraphics[width=0.75\columnwidth]{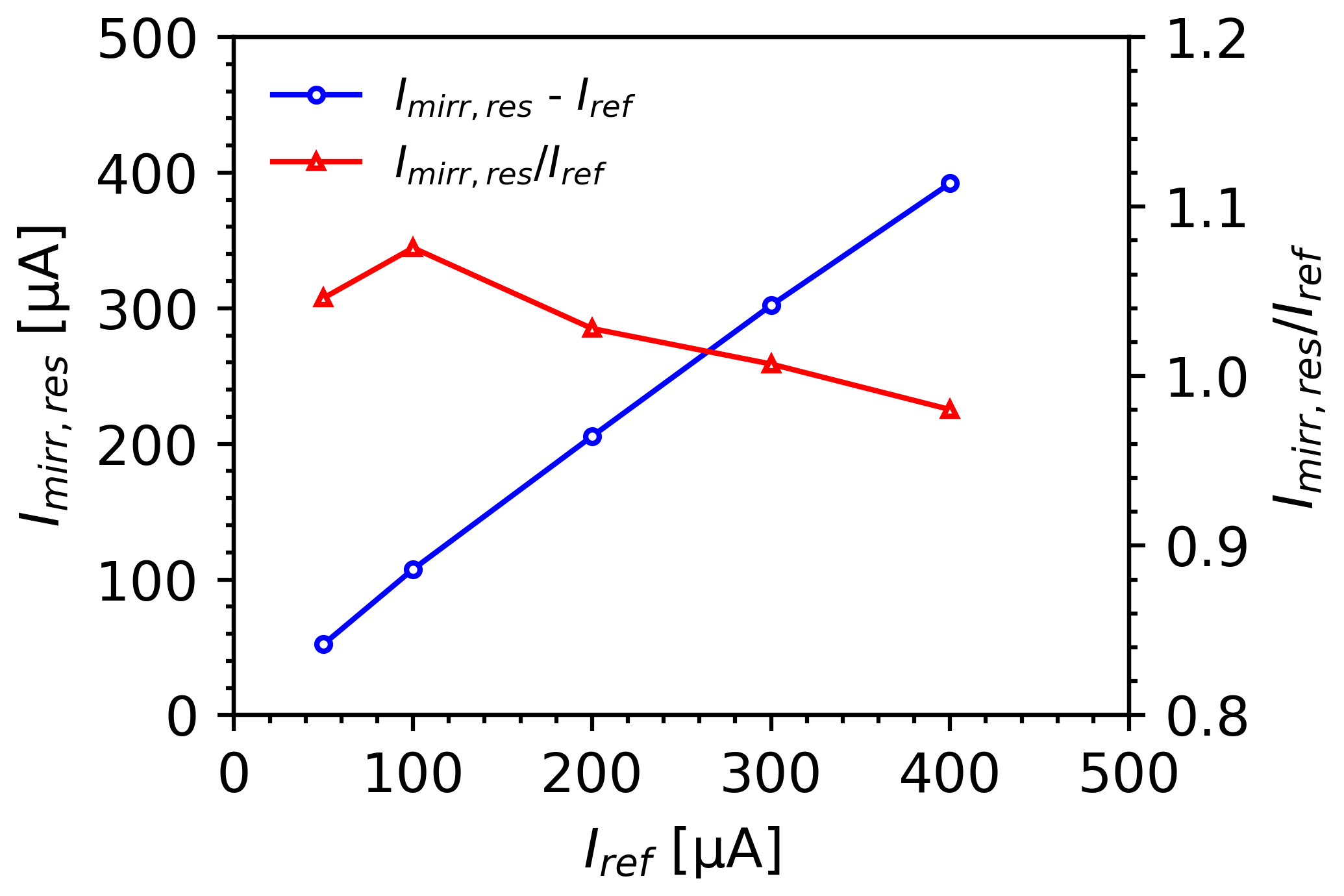}
    \caption{$I_{\mathrm{mirr,res}}$ - $I_{\mathrm{ref}}$ relation and corresponding mirror factors with $I_{\mathrm{ref}}$ ranging from 50\,\textmu A to 400\,\textmu A under fixed $V_{\mathrm{dd,res}}$ of 5\,V.}
    \label{fig:tran RES analyse fac}
\end{figure}

The mirror factor is firstly characterized and calculated in Fig.~\ref{fig:tran RES analyse fac}. Current pulses’ waveforms comparable with SET branch are observable. The waveform of the generated current pulse at each DC reference value match the chop pulse in pulse rising as well as falling edge and duration time. The average pulse amplitude matches its reference value. However, there is a current overshoot observed at the rising edge of the generated pulse. This could be related to the parasitic capacitance. Further, the $I_{\mathrm{mirr,res}}$ - $I_{\mathrm{ref}}$ relation curve as well as mirror factor are plotted and analyzed. As shown in Fig.~\ref{fig:tran RES analyse fac}, $I_{\mathrm{mirr,res}}$ follows $I_{\mathrm{ref}}$ linearly same trend as observed in DC measurement. There is a deviation of the mirror factor observed in the low current range with maximum 7\% at 100\,\textmu A. Overall, the RESET branch is able to generate current pulses across the targeting operation range as a good base for providing enough power to reset \ac{RRAM} in the low current range. \par    

\begin{figure}
    \centering
    \includegraphics[width=0.75\columnwidth]{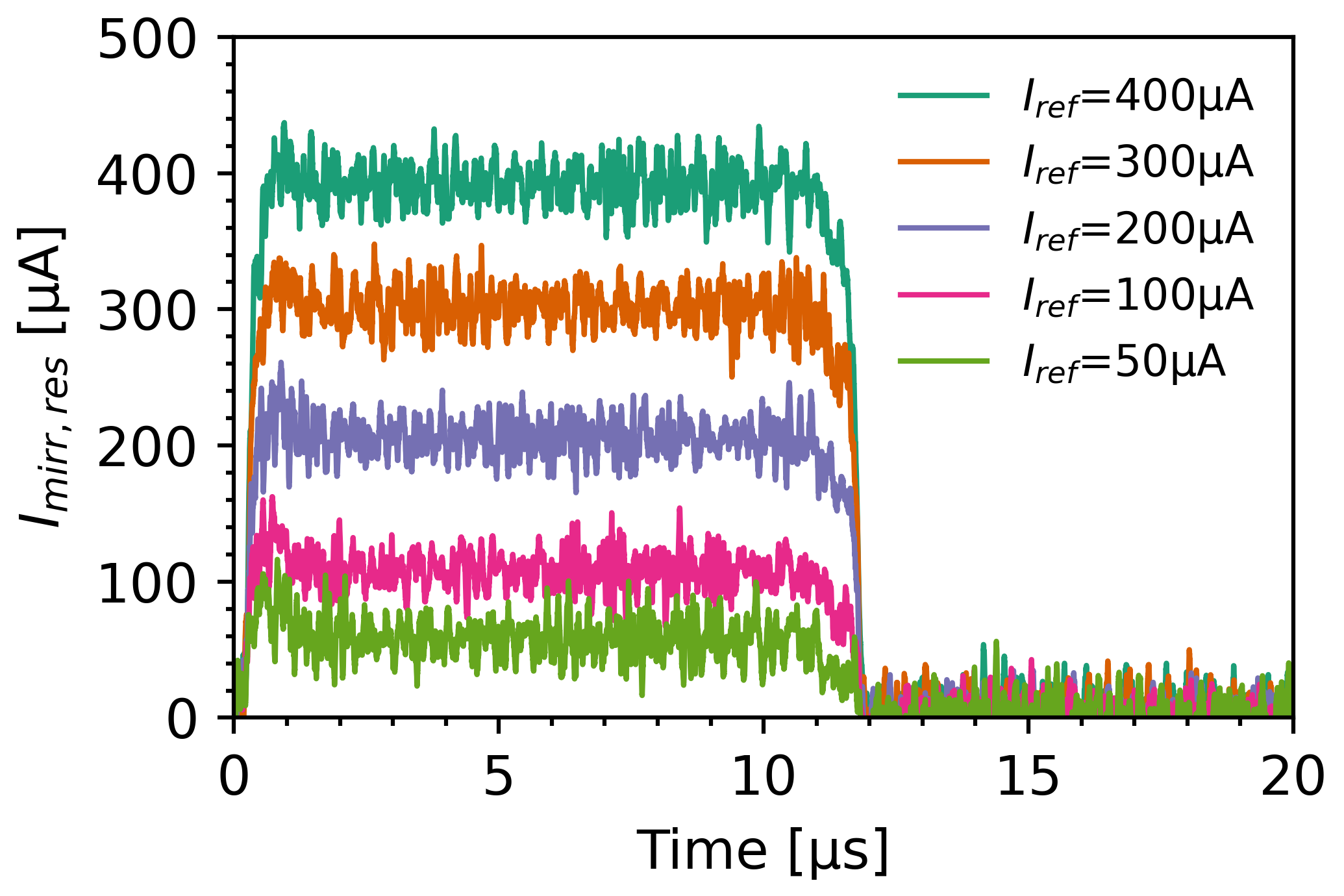}
    \caption{The average of 50 mirrored current pulses with $I_{\mathrm{ref}}$ ranging from 50\,\textmu A to 400\,\textmu A measured from transient measurements on the RESET branch.}
    \label{fig:tran RES analyse pul}
\end{figure}

\begin{figure}
    \centering
    \includegraphics[width=0.75\columnwidth]{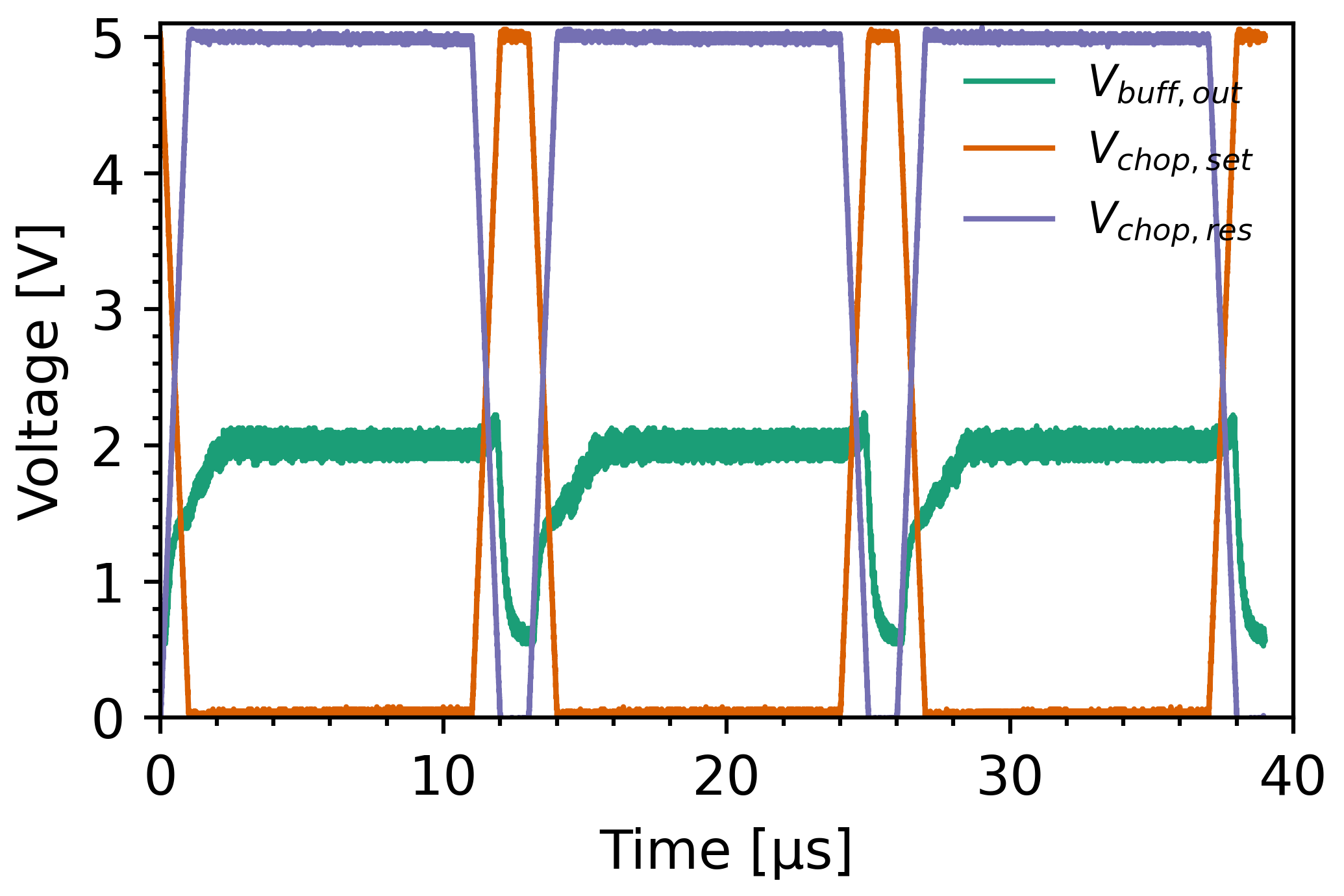}
    \caption{The buffer read out voltage with applying both $V_{\mathrm{chop,set}}$ and $V_{\mathrm{chop,res}}$ synchronously. The delay time between pulses is set to 1\,\textmu s to avoid the parasitic capacitance introduced by floating bottom electrode.}
    \label{fig:tran RRAM branch}
\end{figure}

As for \ac{RRAM} branch, due to the lack of the device, the characterization of this branch mainly relies on the voltage buffer. By measuring the voltage at the input node of the buffer, we can briefly validate the sub branch M2/M1 as well as voltage buffer. To read the input node voltage properly, both M1 and M2 need to be operated in saturation, where both SET and RESET branches are required to be activated. In our case, both $V_{\mathrm{dd,set}}$ as well as $V_{\mathrm{dd,res}}$ are set to 5\,V. Reference DC current source in both branches are programmed to 100\,\textmu A. By activating chop transistors M5 and M10 at same time, two branches carry the generated current pulses to the \ac{RRAM} branch, which leads to the self-regulation of M1 and M2 with proper gate voltage as well as drain voltage. Furthermore, the voltage buffer is activated with $V_{\mathrm{tail}}$ at 1\,V and $V_{\mathrm{dd}}$ 5\,V and the input node voltage change ($V_{\mathrm{D,M1}}$ and $V_{\mathrm{D,M1}}$) is read from buffer output connected to the oscilloscope shown in Fig.~\ref{fig:tran RRAM branch}. It can be observed that read-out voltage $V_{\mathrm{buff,out}}$ increases with applying chop pulses. It researches a stable read-out value around 2\,V within the duration of chop pulses. Later read-out voltage decreases gradually with falling of chop pulses. It needs to be noticed that the read-out voltage falls down to 0.6\,V. This can be addressed to the parasitic capacitance in the branch M2/M1 introduced by the metal pad for \ac{RRAM} bottom electrode deposition. The measured $V_{\mathrm{buff,out}}$ reflects the conductivity of \ac{RRAM} branch and functionality of designed voltage buffer. Moreover, since \ac{NMOS} M1 and \ac{PMOS} M2 in series, to carry same amount current in saturation, the overdrive of M2 should be around 1.4 – 1.7 times larger than M1, which match our measurement result, where the absolute value of M2’s drain source voltage is 3\,V.\par

\section{Discussion}

The proposed \ac{2M1R1B} structure is used to execute current programming of \ac{RRAM}. From the device engineering point of view, the measurement results discussed previously is adequate for our purposes. From DC measurements, SET branch is able to mirror current comparable to the reference value with less than 3\% deviation. This accurate mirroring current will be further copied to \ac{RRAM} by mirror M0:M1 to program the device in current mode without any extra current compliance implementation. The transient measurement's result indicates that the generated current pulse matches the chop pulse well in pulse's waveform. Moreover, the generated current pulse increases gradually at the end of its rising edge rather than overshoot spike caused by parasitic capacitance in \ac{1T1R} structure under the voltage mode. All these features indicate that our proposed \ac{2M1R1B} circuit has a well control of the generated current pulse. As a result, the SET programming of \ac{RRAM} can benefit from this well defined current pulse in better control of filament dynamic. During the SET process, oxygen vacancies' drift under the electrical filed contributes to the ionic current of the whole SET current. With a accurate control of SET current, it would be possible to control the filament evolution during SET precisely and increase the switching uniformity. Furthermore, the accurate mirroring current amplitude brings the flexibility to do multi-level programming within a low current range rather than carefully choosing gate voltage for the transistor as current compliance in \ac{1T1R} structure. With the symmetric design of the SET branch, the RESET branch is also able to generate current pulses within our targeting current range, which brings the possibility to explore the RESET process of \ac{RRAM} with current programming. Additionally, our proposed \ac{2M1R1B} circuit is able to read-out the conductance of the \ac{RRAM} cell with voltage sensing mode by measuring the voltage drop over the cell with constant read current pulse through the voltage buffer. When extended to array level, from the crossbar array programming point of view, the voltage sensing mode used in our approach can lower the power consumption during \ac{VMM} by applying small constant read current. And the sensed voltage cross the cell can be further accumulated and directly processed by the \ac{ADC}, which could prevent the large consumption of power and area peripheral circuits to sink large current while clamping voltage\cite{wan2022compute}.\par 

From the circuit design point of view, \ac{2M1R1B} circuit can be further optimized. As introduced in Fig.~\ref{fig:wafer and circuit}(b), current mirrors in \ac{RRAM} branch are implemented with basic current mirror topography with same \ac{W/L} ratio of MOSFETs to get 1:1 mirror factor. With neglect of channel-length modulation and current mirror working in the saturation, the copied current has no dependence on $V_{\mathrm{ds}}$ and it is only influenced by the \ac{W/L} ratio of implemented MOSFETs as it can be seen from the following equation:

\begin{equation}
    \label{eq:qat:mirror ratio ideal}
    \begin{aligned}
    \frac{I_{D1}}{I_{D0}}
    &=
    \frac{
    \frac{1}{2}\mu_n C_{\mathrm{ox}}
    \left(\frac{W_1}{L_1}\right)
    \left(V_{GS0}-V_{TH}\right)^2
    }{
    \frac{1}{2}\mu_n C_{\mathrm{ox}}
    \left(\frac{W_0}{L_0}\right)
    \left(V_{GS1}-V_{TH}\right)^2}
    =
    \frac{\left(\frac{W_1}{L_1}\right)}{\left(\frac{W_0}{L_0}\right)}  \\[6pt]
    \end{aligned}
\end{equation}

where \textmu$\mathrm{_n}$ is the mobility of charge carriers, $C_{\mathrm{OX}}$ is the gate-oxide capacitance per unit area, $V_{\mathrm{TH}}$ is the threshold voltage, $W_{\mathrm{0}}$ and $W_{\mathrm{1}}$ refer to the gate width of M0 and M1, and $L_{\mathrm{0}}$ as well as $L_{\mathrm{1}}$ refer to the gate length of M0 and M1.
As a result, the reference current pulse can be well mirrored to \ac{RRAM}. However, in practice, channel-length modulation of MOSFET causes errors in copying current as it can be seen from the following equation: 

\begin{equation}
\label{eq:qat:mirror ratio with CLM}
    \begin{aligned}
    \frac{I_{D1}}{I_{D0}}
    &=
    \frac{
    \frac{1}{2}\mu_n C_{\mathrm{ox}}
    \left(\frac{W_1}{L_1}\right)
    \left(V_{GS1}-V_{TH}\right)^2
    \left(1+\lambda V_{DS1}\right)
    }{
    \frac{1}{2}\mu_n C_{\mathrm{ox}}
    \left(\frac{W_0}{L_0}\right)
    \left(V_{GS0}-V_{TH}\right)^2
    \left(1+\lambda V_{DS0}\right)
    } \\[8pt]
    &=
    \frac{W_1 L_0}{W_0 L_1}
    \cdot
    \frac{1+\lambda V_{DS1}}{1+\lambda V_{DS0}}
    \end{aligned}
\end{equation}
where $\lambda$ is the cahnnel-length modulation coefficient.
While $V_{\mathrm{DS0}}$=$V_{\mathrm{GS0}}$=$V_{\mathrm{GS1}}$, $V_{\mathrm{DS1}}$ may not be equal to $V_{\mathrm{GS1}}$ because of the circuity fed by M1, which is the \ac{RRAM} cell. During the SET process, the voltage divider between \ac{RRAM} and transistor is changed, which leads to a variation of $V_{\mathrm{ds}}$ of transistor. Moreover, due to device to device variability, $V_{\mathrm{ds}}$ mismatch is more difficult to minimize. In order to minimize the influence of channel-length modulation, several circuit topologies can be employed. The initial approach involves the implementation of a cascode topology, which serves to shield the current source and thereby mitigate the voltage variation across it \cite{razavidesign}. In practice, as a transform of basic cascode mirror, the configuration "low-voltage cascode" is widely used to reduce voltage supply, which is common issue when designing with cascode structure. Comparing to the basic current mirror, cascode or even low-voltage cascode current mirror can minimize the voltage mismatch.\par 

\section{Conclusion}

In this work, we proposed and designed a circuit to program \ac{RRAM} with current pulses. Thanks to the current mirror topology, defined current can be copied to \ac{RRAM} branch and hence SET process is more controlled and less influenced by the overshoot caused by parasitic capacitance from transistor. By characterizing the circuit in silicon, we demonstrated that our current generator circuit is adequate to execute current programming of \ac{RRAM}.

\begin{acknowledgments}
This work was supported by the European Research Council (ERC) through the European’s Union Horizon Europe Research and Innovation Programme under Grant Agreement No 101042585. Views and opinions expressed are however those of the authors only and do not necessarily reflect those of the European Union or the European Research Council. Neither the European Union nor the granting authority can be held responsible for them. The University of Groningen would like to acknowledge the financial support of the CogniGron research center and the Ubbo Emmius Fund.
\end{acknowledgments}

\section*{Data Availability Statement}

The data that support the findings of this study are available from the corresponding author upon reasonable request.

\bibliography{Reference}

\end{document}